\documentstyle[preprint,aps]{revtex}
\tighten
\draft
\widetext
\preprint{CLNS 00/1672}
\def\simgreat{\mathbin{\lower 3pt\hbox
   {$\rlap{\raise 5pt\hbox{$\char'076$}}\mathchar"7218$}}}

\def\b2{\beta_2}

\def\a2{\alpha_2}

\def\be{\begin{equation}}
\def\ee{\end{equation}}
\def\baray{\begin{eqnarray}}
\def\earay{\end{eqnarray}}

\begin{document}
\title{Cosmology in the Randall-Sundrum Brane World Scenario}
\author{Horace Stoica\footnote{fhs3@mail.lns.cornell.edu}, 
S.-H. Henry Tye\footnote{tye@mail.lns.cornell.edu} and 
Ira Wasserman\footnote{ira@spacenet.tn.cornell.edu}}

\address{Laboratory for Nuclear Studies and Center for Radiophysics 
and Space Research\\
Cornell University \\
Ithaca, NY 14853}

\medskip

\date{\today}
\maketitle

\begin{abstract}
The cosmology of the Randall-Sundrum scenario for a
positive tension brane in a 5-D Universe with localized
gravity has been studied extensively recently. Here we extend it 
to more general situations. We consider the time-dependent situation
where the two sides of the brane are different AdS/Schwarzschild spaces. 
We show that the expansion rate in these models during inflation could
be larger than in brane worlds with compactified extra dimensions of
fixed size. The enhanced expansion rate could lead to the production
of density perturbations of substantially larger amplitude.

\end{abstract}

\vfill\eject

Recently Randall and Sundrum \cite{RS1,RS2} presented a static
solution to the $5$-D (classical) Einstein equations in which
spacetime is flat on a 3-brane with positive tension provided that the
bulk has an appropriate negative cosmological constant. Even if the fifth
dimension is uncompactified, standard $4$-D gravity (specifically,
Newton's force law) is reproduced on the brane. In contrast to the
compactified case \cite{ADD}, this follows because the near-brane 
geometry traps the massless graviton. The extension of their static 
solution to time-dependent solutions and their cosmological properties 
have been extensively studied
 \cite{kaloper,rscos,others,Kraus,Ida,Eanna,bine2,Sasaki,gubser}.
In this paper, we extend further the time-dependent solution to more 
general situations. We shall consider only Friedman-Robertson-Walker 
(FRW) solutions, so the result can be expressed in terms of the Hubble 
constant.
We shall use two approaches to study the problem. One is to solve the 
Einstein equation straightforwardly and obtain the Hubble constant on 
the brane. The bulk properties will be encoded into the behavior 
of the Hubble constant. We shall employ the notation and set-up of 
Ref \cite{binetruy} . The other approach starts with the solution of the 
bulk, and the brane is incorporated using the Israel junction condition
 \cite{Israel,guth} as first used in this context in Ref \cite{Kraus}. 
Here we consider the two sides of the brane to be described by two different
Anti-deSitter(AdS), or  AdS-Schwarzschild (AdSS) spaces,
with different cosmological constants and different Newton constants.
These cosmological scenarios, including the simplest generalizations
of the original Randall-Sundrum brane world that incorporate matter on
the brane \cite{kaloper,rscos,others,Kraus,Ida,Eanna,bine2,Sasaki,gubser},
all can expand much more rapidly at early times than conventional models
based only on 4-D gravity. An important consequence is that the 
amplitude of primordial density fluctuations produced during inflation
could be substantially larger in these scenarios than in brane world
models based on compactified extra dimensions of fixed
size \cite{ADD,dvali,ftw}.

We shall consider a $5$-dimensional Anti-deSitter spacetime with a 
positive-tension $3$-brane sitting inside. 
We are interested in the cosmological solutions with a metric of the
form 
\be
ds^2=-n^2\left( \tau ,y\right) d\tau ^2+a^2\left( \tau ,y\right) \gamma
_{ij}dx^idx^j+b^2\left( \tau ,y\right) dy^2.
\ee
The Einstein equation 
$G_{AB}=\kappa ^2T_{AB}=8{\pi }G T_{AB}$
can be written as\cite{bine2}
\baray
&& G_{00}=3 \left[ \frac{ \dot a}{a} \left( \frac{ \dot a}{a} + 
\frac{ \dot b}b \right) - \frac{n^2}{b^2} \left( \frac{a^{ \prime \prime }}a + 
\frac{a^{ \prime }}a \left( \frac{a^{ \prime}}a- \frac{b \prime}b \right) 
\right) + k \frac{n^2}{b^2} \right] \\
&& G_{ij}= \frac{a^2}{b^2}\gamma_{ij} \left[\frac{a^{ \prime}}a \left( 
\frac{a^{ \prime}}a+2\frac{n^{ \prime}}n\right)-\frac{b^{\prime}}b\left(
\frac{n^{\prime}}n+2\frac{a^{ \prime}}a\right)+2\frac{a^{\prime \prime}}a+
\frac{n^{\prime \prime}}n\right] \nonumber \\ && +\frac{a^2}{n^2}\gamma_{ij}\left[
\frac{\dot a}a\left(-\frac{\dot a}a+2\frac{\dot n}n\right)-2\frac{\ddot a}a
+\frac{\dot b}b\left(-2\frac{\dot a}a+\frac{\dot n}n \right)-\frac{\ddot b}b 
\right] -k\gamma_{ij} \\
&& G_{05}=3\left( \frac{n^{\prime}}n\frac{\dot a}a +\frac{a^{\prime}}a\frac
{\dot b}b - \frac{\dot a^{\prime}}a \right) \\
&& G_{55}=3\left[\frac{a^{\prime}}a\left(\frac{a^{\prime}}a+
\frac{n^{\prime}}n \right)-\frac{b^2}{n^2}\left(\frac{\dot a}a\left(\frac{
\dot a}a-\frac{\dot n}n \right)+\frac{\ddot a}a\right)-k\frac{b^2}{a^2}\right]
\earay
where $\gamma_{ij}$ is a maximally symmetric $3$-dimensional metric, and 
$k=-1,0,1$ parametrizes the spatial curvature. The $3$-brane is placed at $y=0$. 
On the two sides of the brane $\left( y<0 \right) $ and $\left( y>0 \right) $ are 
two different AdS spaces. The stress-energy-momentum tensor has a bulk and a brane 
component:
\be
T_B^A=T_B^A\mid _{bulk}+T_B^A\mid _{brane}.
\ee
The $T_B^A\mid _{brane}$ component corresponds to the matter on the brane
placed at $y=0$. Since the brane is assumed to be homogeneous and isotropic,
this component takes the form 
\be
T_B^A\mid _{brane}=\frac{\delta \left( y\right) }bdiag\left( -\rho
,p,p,p,0\right).
\label{onthebrane}
\ee
The matter density on the brane is composed of a cosmological constant term,
ordinary matter, radiation, etc. $T_B^A\mid _{bulk}$ is the
energy-momentum tensor for the bulk matter, which consists of a non-zero
cosmological constant (assumed to be different on the different sides of the
brane) plus a black hole term, is
\be
T_B^A\mid _{bulk}=diag\left( \Lambda _i,\Lambda _i,\Lambda _i,\Lambda
_i,\Lambda _i\right) 
\ee
where $i=+,-$ i.e. $i=-$ for $y<0$ and $i=+$ for $y>0$. 

Following Ref\cite{binetruy},
it is convenient to define 
$\left[ f\right] =f\left( 0_{+}\right) -f\left( 0_{-}\right)$
to be the {\em jump} component for a given function $f$ at $y=0$,
and 
$\left\{ f\right\} =\left(f^{\prime }\left( 0_{+}\right) +f^{\prime }\left(
0_{-}\right) \right)/2$
to be its {\em average} component at $y=0$.
The functions $n$, $a$, $b$, are continuous at the brane, but their
derivatives are discontinuous, so the second derivatives will be of the form
\be
f^{\prime \prime }=f^{\prime \prime }\mid _{\left( y\neq 0\right) }+\left[
f^{\prime }\right] \delta \left( y\right).
\ee
Using this notation, the jump part of the Einstein equation for the 
$(00)$, the $(ij)$ and the $(55)$ components can be written as (the 
subscript $0$ indicates that the functions are evaluated at $y=0$)
\baray
\label{jump_00}
&& \frac{2\left\{ a^{\prime }\right\} }{a_0b_0}\frac{\left[ a^{\prime }\right] 
}{a_0b_0}-\frac{\left[ a^{\prime }\right] }{\left( a_0b_0\right) }\frac{
\left\{ b^{\prime }\right\} }{b_0^2}-\frac{\left\{ a^{\prime }\right\} }{
\left( a_0b_0\right) }\frac{\left[ b^{\prime }\right] }{b_0^2}=\frac{\kappa
^2}3\left(\Lambda_+-\Lambda_-\right)  \\
\label{jump_ij}
&& \frac{2\left\{ a^{\prime }\right\} }{a_0b_0}\frac{\left[ a^{\prime }\right] 
}{a_0b_0}+2\left( \frac{\left[ a^{\prime }\right] }{a_0b_0}\frac{\left\{
n^{\prime }\right\} }{n_0b_0}+\frac{\left\{ a^{\prime }\right\} }{a_0b_0}
\frac{\left[ n^{\prime }\right] }{n_0b_0}\right) -\left( \frac{\left[
b^{\prime }\right] }{n_0b_0}\frac{\left\{ n^{\prime }\right\} }{b_0^2}+\frac{
\left\{ b^{\prime }\right\} }{b_0^2}\frac{\left[ n^{\prime }\right] }{n_0b_0}
\right) \nonumber \\ 
&& - 2\left( \frac{\left[ a^{\prime }\right] }{a_0b_0}\frac{
\left\{ b^{\prime }\right\} }{b_0^2}+\frac{\left\{ a^{\prime }\right\} }{
a_0b_0}\frac{\left[ b^{\prime }\right] }{b_0^2}\right) =\kappa ^2\left(\Lambda_+-\Lambda_-\right) \\
\label{jump_55}
&& \frac{2\left\{ a^{\prime }\right\} }{a_0b_0}\frac{\left[ a^{\prime }\right] 
}{a_0b_0}+\frac{\left[ a^{\prime }\right] }{a_0b_0}\frac{\left\{ n^{\prime
}\right\} }{n_0b_0}+\frac{\left\{ a^{\prime }\right\} }{a_0b_0}\frac{\left[
n^{\prime }\right] }{n_0b_0}=\frac{\kappa ^2}3\left(\Lambda_+-\Lambda_-\right) 
\earay
while the $\delta $-function part of the Einstein equation for the
$(00)$ and the $(ij)$ components can be written as

\be
\label{delta}
\frac{\left[ a^{\prime }\right] }{a_0b_0}=-\frac{\kappa ^2}3\rho \; ,\;
\frac{\left[ n^{\prime }\right] }{n_0b_0}=\frac{\kappa ^2}3\left( 2\rho
+3p\right).
\ee
The remaining (the {\em average}) part of the $(55)$ component of the 
Einstein equation is given by

\baray
\label{G_55_avg}
&& \frac 1{n_0^2}\left[ \frac{\dot a_0}{a_0}\left( \frac{\dot a_0}{a_0}-\frac{
\dot n_0}{n_0}\right) +\frac{\ddot a_0}{a_0}\right] =\frac 14\left( \frac{
\left[ a^{\prime }\right] }{a_0b_0}\right) ^2+\left( \frac{\left\{ a^{\prime
}\right\} }{a_0b_0}\right) ^2 \nonumber \\ 
&& +\frac 14\frac{\left[ a^{\prime }\right] }{
a_0b_0}\frac{\left[ n^{\prime }\right] }{n_0b_0}+\frac{\left\{ a^{\prime
}\right\} }{a_0b_0}\frac{\left\{ n^{\prime }\right\} }{n_0b_0}-\frac{\kappa
^2}3\frac{\Lambda_++\Lambda_-}2 -\frac{k}{a_0^2}.
\earay

Using the above equations and switching to the
proper time of the brane $t$ defined by $d\tau =n\left( \tau ,0\right) dt$,
the function $a_0(t)=a\left( t,0\right) $
describes the evolution of our four-dimensional universe, and will be
denoted by $R\left( t\right) $. Then the {\em average} component of 
$G_{55}=\kappa ^2T_{55}$ gives the Hubble constant 
$H={\dot{a_0}}/{a_0}={\dot{R}}/{R}$ of the brane. The steps to follow are the 
same as in Ref\cite{rscos,Eanna}. After going to the proper time of the 
brane, the LHS of Eq(\ref{G_55_avg}) becomes 
\be
\frac1R\frac{d^2R}{dt^2}+\frac1{R^2}\left(\frac{dR}{dt}\right)^2=
\frac{\dot R}R\frac{d\dot R}{dR}+\frac1{R^2}\left(\frac{dR}{dt}\right)^2=
\frac{R}2\left(\frac{dH^2}{dR}\right)+2H^2=
\frac 1{2R^3}\frac d{dR}\left( H^2R^4\right).
\ee
The Hubble constant of the brane can be obtained by integrating 
the equation  
\baray
\label{hub}
&& \frac 1{2R^3}\frac d{dR}\left( H^2R^4\right) =-\frac{\kappa ^4}{36}\rho
\left( \rho +3p\right) -\frac{\kappa ^2}3\frac{\Lambda_++\Lambda_-}2 
-\frac{k}{R^2}  +\left( \frac{\left\{ a^{\prime }\right\} }{
a_0b_0}\right) ^2\left[ 1+\frac{3p}\rho \right] \nonumber \\
&& +\left( \frac{\left\{a^{\prime }\right\} }{a_0b_0}\right) 
\frac{\left(\Lambda_+-\Lambda_-\right) }\rho.
\earay

The function $\left\{ a^{\prime }\right\} / a_0b_0$ is determined by 
the properties of the bulk. That is, this brane function encodes  
informations coming from the bulk.
If the two sides of the brane are identical AdS spaces,
it becomes zero, as required by the $Z_2$ symmetry. 
(All functions depend on $y\,$ only through $\left| y\right| $.) 
The other functions 
$\left\{ b^{\prime }\right\} / b_0^2$, $\left\{ n^{\prime }\right\} / n_0b_0$, 
$\left[ b^{\prime}\right] / n_0b_0$, can be determined in terms of $\left\{ a^{\prime
}\right\} / a_0b_0$. Since $\left[a^{\prime}\right]/a_0b_0$ and $\left[n^{\prime}\right]/n_0b_0$
are already determined by Eq(\ref{delta}), then Eq (\ref{jump_00}), 
(\ref{jump_ij}) and (\ref{jump_55}) allow us to find the three other unknown functions.
We may consider expanding $\left\{ a^{\prime }\right\} / a_0b_0$ 
in a series of even powers of $1/R$:
\be
\label{coef}
\frac{\left\{ a^{\prime }\right\} }{a_0b_0}=c_0+\frac{c_1}{R^2}+\frac{c_2}{
R^4}+\frac{c_3}{R^6}\ldots  
\ee
We shall see some examples where these coefficients have specific physical 
interpretations.

An alternative way to obtain the Hubble constant has been used in 
Ref\cite{Kraus}. The bulk is made of two pieces of
five-dimensional AdSS space-time separated by the brane.
The five-dimensional action is
\be
\label{action}
S=\frac 1{16\pi G}\int_Md^5x\sqrt{-g}\left( R+\frac{12}{l^2}\right) +\frac 1{
8\pi G}\int_{\partial M}d^4x\sqrt{-\gamma }K 
\ee
The cosmological constant used here and in the Ref\cite{Kraus}, $1/l^2$, is 
related to the cosmological constant $\Lambda$ used in Ref\cite{RS1,RS2,Eanna}
by $1/l^2=\kappa^2\Lambda/6$. The brane separating the two AdSS spaces is 
described by the equation $r=R \left( t \right)$. The extrinsic curvature of 
the brane, $K ^{ab}=\nabla^an^b $ can be written as
$K_{\mu\nu} = n^{c} \partial _c \gamma _{\mu\nu}/2$
where $\gamma _{\mu\nu}$ is the induced metric on the brane, and $n^{a}$ is the 
unit normal of the brane. The indices $a,b$ cover the 5D space-time,
the indices $\mu,\nu$ cover the 4D space-time, and the indices $i,j$ cover the 
space coordinates of the brane. 
The junction condition at the brane as required by the Einstein equations is
\be
\frac{K_{ij}^{+}}{G_+}-\frac{K_{ij}^{-}}{G_-}=-8\pi \left( T_{ij}-\frac 
13T \gamma_{ij}\right)
\ee
where $T_{\mu\nu}$ is the energy-momentum tensor of matter on
the brane, and $T$ is its trace.
Here we generalize Eq(\ref{action}) 
to allow the possibility that the Newton's constants on the two sides are different.
This scenario can happen in string theory, when both the cosmological 
constant and the Newton's constant of an AdS space are related
to the number of D-branes present. So we may visualize the situation where 
the two AdS bulk spaces separated by the brane may have different
Newton's constants and cosmological constants.
We may also entertain the possibility that the cosmological constants on 
the two sides are different, a situation that can arise when the brane is a 
thin-wall approximation of a domain wall \cite{Soleng}.
Note that the solution of the jump
condition yields a solution to the $5$-D Einstein equation, in contrast to
the case where the brane is treated as a probe \cite{vatche}.

If the bulk metric has the form
\be
ds^2 = -f \left( r \right) dt^2+r^2 d \Sigma _k ^2
+f ^{-1} \left( r \right) dr^2
\ee
then the velocity vector of the brane $u^{\mu}$, which satisfies $u^{ \mu } 
u_{\mu} = -1$ and $n^{\mu} u_{\mu} = 0$ is given by 
$u^t = \left( f + \dot R^2 \right) ^{ \frac{1}{2} } f ^{-1} $ and
$u^r = \dot R $, where
$\dot R$ being the derivative with respect to the proper time $ \tau $.
Up to a sign, the unit normal to the brane is given by
$n^t=-f ^{-1} \dot R $ and $\ n^r=-\left( f+\dot R^2\right)^{\frac{1}{2}}$.
The minus sign is due to the fact that the coordinate $r$ is decreasing in 
the direction $n^{\mu}$. With these components, the spatial components of the 
extrinsic curvature on the two sides of the brane are
\be
K_{ij}^-=\frac{\left( f_-+\dot R^2\right)^{\frac{1}{2}}}{R}\gamma_{ij}\ , \ 
K_{ij}^+=-\frac{\left( f_++\dot R^2\right) ^{\frac{1}{2}}}{R}\gamma_{ij}
\ee
where the relative signs of $K_{ij}^-$ and $K_{ij}^+$ follow from the 
definition of the unit normal  $n^{\mu}$.

For a matter tensor on the brane of the form given in Eq(\ref{onthebrane}), 
$T_{ij}-{1\over 3}T\gamma_{ij}=(\sigma+\rho_m)/3$, and
the equation describing the 
evolution of the brane (3-brane) becomes
\be
\frac{\left( f_-+\dot{R}^2\right) ^{\frac 12}}{G_-}+\frac{\left( f_++\dot{R}
^2\right) ^{\frac 12}}{G_+}=\frac{8\pi \left(\sigma+\rho _m \right) }3R.
\ee
Using the above 
equation and the notation $\tilde \lambda=8\pi \left(\sigma+\rho_m \right)/3$, 
the Hubble constant is found to be:
\baray
&& H^2=\left(\frac{\dot R}R\right)^2=\frac{\tilde \lambda^2\left(G_+^2+G_-^2\right)
G_+^2G_-^2}{\left(G_+^2-G_-^2\right)^2}+
\frac{\left(f_+G_-^2-f_-G_+^2\right)}{R^2\left(G_+^2-G_-^2\right)}\nonumber \\
&& -\frac{2G_+^2G_-^2}{\left(G_+^2-G_-^2\right)^2}\left[
\tilde \lambda^4G_+^2G_-^2+\frac{\tilde \lambda^2}{R^2}\left(f_+-f_-\right)
\left(G_+^2-G_-^2\right)\right]^{\frac12}
\earay
For an explicit example, we can consider the metric given in
Ref\cite{Birm},
\be
f_{\pm}=k+\frac{R^2}{l^2_{\pm}}-\frac{\mu_{\pm}}{R^2}.
\ee
This reduces to the case considered in Ref\cite{Kraus} if we set
$G_+=G_-=G$ and $l_+=l_-=l$. The existence of a horizon at 
$r_{h_i}^2=\frac{l_i^2}2\left( -k\pm \sqrt{k^2+4\mu _i/l_i^2}\right) \,$ 
imposes restrictions on the values of $\mu
_i\,$ depending on the value of $k$. For $k=+1$, a positive value of $
r_{h_i}^2$ imposes $\mu _i>0$ , while for $k=-1$, the condition is $\mu
_i>-l_i^2/4$.

We may also consider a more general metric. Motivated by the generalized 
AdSS solution of Ref \cite{Cvetic} , we can 
consider a metric of the form
\be
ds^2=-\omega_i^{-2}f_idt^2+\omega_i\left( f^{-1}_idr^2+r^2d\Omega_{3,k} \right)
\ee
where $f=k-\frac{\mu }{r^2}+\frac{r^2}{l^2}\omega^3$. 
In Ref\cite{Cvetic}, $\omega^3= H_1H_2H_3$ with $H_I= 1 +\frac{q_I}{r^2}$,
where $q_I$ are charges. 
For $\omega ^3=1$ and $G_+=G_-$, it reduces to the metric of Ref \cite{Birm}.
In general $\omega$ is a function of $r$. In this case, we have
\be
u^r=\dot{R},\; u^t=\left( \dot{R}^2+\frac f\omega \right)^{1/2}\frac{\omega ^3}f,\; 
n^t=-\frac{\omega ^{3/2}}f\dot{R},\; n^r=-\left( \frac f\omega +
\dot{R}^2\right) ^{1/2}.
\ee
It is now straightforward to use the jump condition to obtain an expression 
for the Hubble constant. To match the two approaches we have discussed, let us 
go back to the simple case where $\omega=1$ and  $G_+=G_-$. The equality 
$G_+=G_-=G$ allows us to define $\lambda=G \tilde \lambda$.
We can solve for $\dot R$ and obtain the Hubble constant:
\be
H^2=\left( \frac{\dot{R}}R\right) ^2=\frac{\lambda ^2}4-\frac{f_-+f_+
}{2R^2}+\frac{\left( f_--f_+\right) ^2}{4R^4\lambda ^2} 
\ee
Using the expressions for the functions $f_i$, the Hubble constant becomes 
\baray
\label{H_Kraus}
&& H^2=\frac{\lambda ^2}4-\frac 12\left( \frac 1{l_-^2}+\frac 1{l_+^2}\right) +
\frac 1{4\lambda ^2}\left( \frac 1{l_-^2}-\frac 1{l_+^2}\right) ^2+ 
\frac {k}{R^2} \nonumber \\
&& +\frac 1{R^4}\left\{ \frac{\mu_-+\mu_+}2-\frac{\mu_+-\mu_-}{
2\lambda ^2}\left( \frac 1{l_-^2}-\frac 1{l_+^2}\right) \right\} +
\frac 1{R^8}\frac{\left( \mu_+-\mu_-\right) ^2}{4\lambda ^2}.
\earay

To compare Eq(\ref{hub}) with this expression, we must first integrate
Eq(\ref{hub}), which will generate an integration constant. We see that
the integration constant is
\be
C=\frac{\mu_-+\mu_+}2-\frac{
\mu_+-\mu_-}{2\lambda ^2}\left( \frac 1{l_-^2}-\frac 1{l_+^2}\right) 
\ee
where we identify
\be
\lambda =\frac{\kappa ^2\sigma }3, \;
\frac 1{l_i^2}=\frac{\kappa ^2\Lambda _i}6 
\ee
and $c_0=-\kappa ^2\left( \Lambda_+-\Lambda_-\right)/12, \; c_1=0, \;
c_2=-\left(\mu_+-\mu_-\right)/2\lambda$. The integration constant was obtained 
in Ref\cite{Kraus,Eanna,bine2}, and the $R^{-8}$ term was first obtained in 
Ref\cite{Kraus}. 
The $\mu$ term has been interpreted as due to a $N=4$ 
super-Yang-Mills theory on the brane via the AdS/CFT 
correspondence \cite{gubser}. 
The holographic principle \cite{lenny}
states that information in the
bulk is encoded in the data on the boundary. Intuitively, one may
understand this as a boundary value problem. Evolution of the
boundary to the bulk fixes the properties of the bulk in terms
of the values at the boundary. However, this intuition does not
apply to the brane world scenario when there are different AdSS
spaces on the two sides of the brane.
It is natural to ask if data on the brane
encodes all properties of both bulks, or only a combination of
the bulk information with no chance to dis-entangle them. The
presence of the $(\mu_+-\mu_-) ^2/R^8$ term in addition to the
$(\mu_++\mu_-) ^2/R^4$ term in the Hubble constant equation
allows one to determine both $\mu_+$ and $\mu_-$. That is, there are 
two different conformal field theories on the brane, which couple to 
each other. This suggests that there is enough information on the brane to
determine fully all the bulk properties on each side of the brane.
In this sense, the holographic principle is deep.

Tuning the effective cosmological constant in the brane to zero requires
\be
\frac{\lambda ^2}4-\frac 12\left( \frac 1{l_-^2}+\frac 1{l_+^2}\right) +
\frac 1{4\lambda ^2}\left( {\frac 1{l_-^2}-\frac 1{l_+^2}}\right) ^2=0 
\ee
or using $\lambda =\frac{\kappa ^2\sigma }3$,
\be
\frac{\kappa ^4\sigma ^2}{36}-\frac{\kappa ^2\left( \Lambda_-+\Lambda_+\right) }{12}
+\frac{\left( \Lambda_--\Lambda_+\right) ^2}{16\sigma ^2}=0.
\ee
For $\Lambda_-=\Lambda_+$, this reduces to the Randall-Sundrum case.
Otherwise, the equation has the two solutions
\be
\label{sol}
\kappa ^2\sigma ^2=\frac 32\left( \sqrt{\Lambda_-}\pm \sqrt{\Lambda_+}
\right) ^2.
\ee
the $''+''$ solution being the one found in Ref\cite{shirman}. 
Note that, if we choose either $\Lambda$ to be zero, the two solutions 
for the value of the brane tension merge, and the minimum of $H^2$ is exactly zero.
By choosing $\mu_i=0$ and $\rho_m=0$ in Eq(\ref{H_Kraus}) we obtain
\be
H^2=\frac{\kappa^4\sigma^2}{36}-\frac{\kappa^2\Lambda}{12}+\frac{\Lambda^2}{16\sigma^2}=
\left(\frac{\kappa^2\sigma}{6}-\frac{\Lambda}{4\sigma}\right)^2
\ee
This intriguing property implies that the minimum of the $4$-dimensional 
effective cosmological constant is bounded below by zero. 
 
The relationship between the 5D and 4D Newton constants is
\be
\label{G4-G5}
\frac{l_+}{G_+} +\frac{l_-}{G_-}=\frac2{G_{\left( 4\right) }}.
\ee
This can be obtained by dimensional reduction. 
We may also identify the 4D Newton constant
by adding matter to the brane, expanding for $\rho _m\ll \sigma $ and
identifying the coefficient of $\rho _m $ in Eq(\ref{H_Kraus}). 

This shows that conventional cosmology
can be recovered at large enough values of $R$, where the matter density
is low and the additional terms in Eq(\ref{H_Kraus}) become small.
Choosing the ``$-$'' solution of Eq(\ref{sol}) does not, however, lead to
conventional cosmology at large enough $R$.

It is instructive to consider a simplified version of Eq(\ref{H_Kraus})
to better appreciate its significance. To be concrete, let us assume that
$k=0=\mu_+-\mu_-$, but still allow $l_-\neq l_+$. Then we find that 
\be
H^2={(\lambda^2-\lambda_+^2)(\lambda^2-\lambda_-^2)\over 4\lambda^2}
+{\mu\over R^4},
\label{H_simplified}
\ee
where $\mu=\mu_\pm$ and $\lambda_\pm^2=(1/l_+\pm 1/l_-)^2$; quite
generally, this is to be augmented by an equation governing the evolution
of $\lambda$, which can always be written schematically as
$\dot\lambda=-3H(1+w)\lambda$,
where for mixtures of matter, radiation, and evolving fields,
$-1\leq w\leq 1$, so $\lambda$ decreases with expansion generically.
It is easy to
see that the first term in Eq(\ref{H_simplified}) is negative for
any $\lambda^2<\lambda_+^2$, which means that for such values of
$\lambda$, there are only sensible cosmological models if
$\mu>0$. In such a case, one expects that the Universe expands to 
a maximum $R$ and then recollapses, generically, although it is
conceivable that the parameters could be fine-tuned to avoid recollapse
and attain $\lambda\to\lambda_-$ as $R\to\infty$. However,
it is clear that the $\mu/R^4$ term plays an essential role in the
expansion rate throughout the history of such models, whether they
collapse or expand forever, and consequently
conventional cosmology is not recovered in any limit. Thus,
such models could never reproduce the successes of cosmological
nucleosynthesis theory, for example, and would not yield acceptable
theories of large scale structure, even if expansion to very large
(or infinite) $R$ is possible.

On the other hand, for $\lambda>
\lambda_+$, the Universe always expands, irrespective of the sign of
$\mu$, although, for $\mu<0$ it expands from a minimum value of $R$.
As long as there is some component in $\lambda$ with $w<1/3$, such
as the brane tension or the vacuum energy density of a brane field, the
$\mu/R^4$ tern in Eq(\ref{H_simplified}) becomes progressively
less important as the Universe expands, and conventional cosmology
can be recovered provided that $\lambda\to\lambda_+$ as
$R\to\infty$.

During the inflation era, the matter density, $\rho_m$, is dominated
by the inflaton, which we will take to be a single component scalar field
$\phi$ with some effective potential $V(\phi)$. The inflaton will tend to
roll toward its potential minimum, and, while the field is rolling slowly,
the energy density in the field is dominated by $V(\phi)$, which decreases
with time as the Universe expands, but only slowly if $V(\phi)$ is
flat enough.
If the inflation commences with
$\lambda>1/l_++1/l_-$, then  $H^2$ decreases as
$\phi$ rolls toward the minimum of $V(\phi)$, where $V(\phi)=V_{min}$.
Provided that $V_{min}+\sigma$ is tuned so that $\kappa^2(V_{min}+
\sigma)/3=\lambda_+$, conventional cosmology can be recovered.
\footnote{Note that to the extent that we may think
of $\sigma$ as the tension of a domain wall solution for self-gravitating
supergravity fields off the brane, and $V_{min}$ as arising from standard
model fields that are confined to the brane, the precise cancellation of
the cosmological constant involves cooperation between brane and bulk
physical fields.}

One of the intriguing features of Eq(\ref{H_Kraus}) is that for
$\lambda$ well above $1/l_+^2+1/l_-^2$, the expansion rate grows linearly
with $\lambda$, i.e. $H\approx\lambda/2$ (assuming that the other
terms in Eq(\ref{H_Kraus}) that are proportional to inverse
powers of $R^2$ can be neglected).\footnote{This is also true in
scenarios where the AdS spaces on either side of the brane are
identical.} With $\rho_m\approx V(\phi)$,
this limit applies when $V(\phi)$ is at least a factor of a few
larger than $\sigma_+=\sqrt{3/2}(\sqrt{\Lambda_+}+\sqrt{\Lambda_-})/\kappa$.
The expansion rate during this
epoch, $H\approx\kappa^2V(\phi)/6$, can be much larger than the
expansion rate in conventional inflation for the same value of
$V(\phi)$, which is $H_{conv}=(8\pi V(\phi)/3M_P^2)^{1/2}$,
where $M_P$ is the Planck mass: $H/H_{conv}\approx 
\kappa^2M_P\sqrt{V(\phi)/96\pi}=
\sqrt{V(\phi)/8\sigma_+}(\sqrt{l_+/l_-}+\sqrt{l_-/l_+})$. 
Conceivably, $H\to H_{conv}$ only as $\phi$ settles into its
potential minimum, and during much of inflation
$H\approx\kappa^2V(\phi)/6$ instead. This can happen in a number
of ways. First, if $V(\phi)$ has some characteristic scale $V_0$,
which can occur if the effective potential is flat until it
plummets to its minimum at $V_{min}\ll V_0$ (or $V_{min}=0$), 
and $V_0\gg\sigma_+$, then $H\approx\kappa^2V_0/6$
for most of the inflationary era. Second, if $V(\phi)$ is
an increasing, polynomial function of $\phi$, for example
a simple powerlaw $V(\phi)\sim\phi^n$, then $H=\kappa^2V(\phi)/6$
at sufficiently large values of $\phi$, and the conventional
expansion rate is only achieved when $V(\phi)\lesssim\sigma_+$.
Finally, when either $l_+$ or $l_-$ is much larger than the
other, then the expansion rate can be $H\approx\kappa^2V(\phi)/6$
even when $V(\phi)$ is only a factor of a few larger than 
$\sigma_+$.

Generally speaking, this enhanced expansion rate would have no
consequence for the mean properties of the Universe; reheating
and the end of inflation would be unaffected.  But, the production
of density perturbations from fluctuations in the inflaton field
could be dramatically different than in models for brane world
inflation with compactified extra dimensions of fixed size, where
$H=H_{conv}$ and 
$\delta\rho/\rho$ generally is too small\cite{ADD,dvali,ftw}.
To evaluate the differences, suppose
that $\sigma=\sigma_+$ and $V_{min}=0$. We estimate the magnitude
of the density fluctuations produced from the usual relationship
$(\delta\rho/\rho)_q\sim (V^\prime(\phi)H/\dot\phi^2)_{h.c.}\sim
(H^3/V^\prime)_{h.c.}$, where $V^\prime=\partial V(\phi)/\partial\phi$,
and the subscript ``h.c.'' means that we evaluate $H^3/V^\prime$
when a perturbation of comoving scale $q^{-1}$ crosses the horizon
during inflation \cite{perts}.
Two simple examples will
suffice to illustrate the effect. If $V(\phi)=m^2\phi^2/2$, an
example of ``chaotic inflation'' (e.g. \cite{linde}), then
we estimate $\delta\rho/\rho\sim\kappa m^{3/2}N_q^{5/4}$, which
is to be compared to the value 
$(\delta\rho/\rho)_{conv}\sim (m/M_P)N_q$ for $H=H_{conv}$,
where $N_q$ is the
number of e-foldings of $R(t)$ remaining in inflation
after horizon crossing for a comoving scale $q^{-1}$.
The fluctuations on scales that cross the horizon
while $H\approx\kappa^2V(\phi)/6$ differ from what one would
get for $H=H_{conv}$ by a factor $\sim\kappa m^{1/2}M_P
N_q^{1/4}$. If instead $V(\phi)=V_0[1-\exp(-\phi/m)]$, which
arises in ``brane inflation'' (e.g. \cite{dvali}, \cite{ftw}),
then we estimate $\delta\rho/\rho\sim\kappa^2 V_0N_q/m$,
which differs from the estimate $(\delta\rho/\rho)_{conv}
\sim V_0^{1/2}N_q/M_Pm$ found for $H=H_{conv}$
by a factor $\sim\kappa^2V_0^{1/2}M_P$.
If the characteristic energy scales of these inflationary models
are comparable to $\kappa^{-2/3}$, and $M_P\gg\kappa^{-2/3}$,
then the implied density fluctuations are much larger than for
$H=H_{conv}$ with the same inflaton potentials.
The enhancement in $\delta\rho/\rho$ for these two models is a
consequence of the faster expansion rate during inflation,
so that the $H^3$ factor in our estimate of $\delta\rho/\rho$
is larger than $H_{conv}^3$, but is mitigated
by the evolution of $\phi$, which tends to raise the value
of $V^\prime$ at horizon crossing for these potentials. 
\footnote{There are also potentials for which there is no
effect, and for which the suppression due to $V^\prime$
outweighs the enhancement due to $H^3$. Thus, for 
$V(\phi)=\lambda\phi^4/4$ or $V(\phi)=V_0-\lambda\phi^4/4$.
we estimate $\delta\rho/\rho\sim\lambda^{1/2}N_q^{3/2}$,
just as in conventional inflation, and for $V(\phi)=
m^{4-n}\phi^n/n$ with $n>4$ the density fluctuations
may even be smaller than for $H=H_{conv}$.}
A similar effect is seen in brane world cosmologies in which
the extra dimensions are compactified, provided that the
extra dimensions are smaller, and therefore the 
effective Newton ``constant'' is larger, during inflation 
than today \cite{ftw,gia}. 

We thank Eanna Flanagan and Vatche Sahakian for useful discussions.
The research of S.-H.H.T. is partially
supported by the National Science Foundation.  I.W. gratefully
acknowledges support from NASA.

\end{document}